\documentclass[12pt]{article}

\pdfoutput=1

\usepackage{color}
\usepackage{amsmath}
\usepackage{amsfonts}
\usepackage{amssymb}
\usepackage{caption}
\usepackage{graphicx}
\usepackage{slashed}            
\usepackage{subfig}             
\usepackage{xspace}				
\usepackage{cite}

\usepackage[english]{babel}
\usepackage{fancyhdr}
\usepackage{amsmath}
\usepackage{amssymb}
\usepackage{amsfonts}
\usepackage{psfrag}
\usepackage[applemac]{inputenc}
\usepackage{graphicx}

\def\gsim{\mathrel{\rlap{\lower4pt\hbox{\hskip1pt$\sim$}}
    \raise1pt\hbox{$>$}}}                

\newcommand{\arXiv}[2]{\href{http://arxiv.org/pdf/#1}{{\tt #2/#1}}}
\newcommand{\arXivold}[1]{\href{http://arxiv.org/pdf/#1}{{\tt #1}}}

\newcommand{\TeV}{\,\mathrm{TeV}}
\newcommand{\GeV}{\,\mathrm{GeV}}

\newcommand{\beq}{\begin{eqnarray}}
\newcommand{\eeq}{\end{eqnarray}}

\newcommand{\eq}[1]{Eq.~(\ref{#1})}
\newcommand{\fig}[1]{Fig.~\ref{#1}}

\newcommand{\bag}{\begin{align}}
\newcommand{\eag}{\end{align}}

\newcommand{\keV}{\,\mathrm{keV}}

\newcommand{\vev}[1]{\langle {#1} \rangle}

\newcommand{\Lag}{\mathcal{L}}

\usepackage[hypertexnames=false]{hyperref}		

\begin{document}
\begin{titlepage}

\vskip.5cm

\begin{center} 
{\huge \bf Inflation from Broken Scale Invariance  \\ \vspace*{0.3cm} } 
\end{center}

\begin{center} 
{\bf \  Csaba Cs\'aki$^a$, Nemanja Kaloper$^b$,  Javi Serra$^a$, and John Terning$^b$} 
\end{center}

\begin{center} 

$^{a}$ {\it Department of Physics, LEPP, Cornell University, Ithaca, NY 14853, USA} \\

\vspace*{0.1cm}

$^{b}$ {\it Department of Physics, University of California, Davis, CA 95616} \\

\vspace*{0.1cm}

{\tt  
\href{mailto:csaki@cornell.edu}{csaki@cornell.edu}, 
\href{mailto:csaki@cornell.edu}{kaloper@physics.ucdavis.edu},
 \href{mailto:js993@cornell.edu}{js993@cornell.edu},
 \href{mailto:jterning@gmail.com}{jterning@gmail.com}}

\end{center}

\vglue 0.3truecm

\centerline{\large\bf Abstract}
\begin{quote}
We construct a model of inflation based on a low-energy effective theory of spontaneously broken 
global scale invariance. This provides a shift symmetry that protects the inflaton 
potential from quantum corrections. Since the underlying scale invariance is non-compact, arbitrarily large inflaton field displacements
are readily allowed in the low-energy effective theory. A weak breaking of scale invariance by almost 
marginal operators provides a non-trivial inflaton minimum, which sets and stabilizes 
the final low-energy value of the Planck scale. The underlying scale invariance ensures that the slow-roll approximation
remains valid over large inflaton displacements, and yields a scale invariant spectrum
of perturbations as required by the CMB observations.
 
\end{quote}

\end{titlepage}


\setcounter{equation}{0}
\setcounter{footnote}{0}

\section{Introduction}
\setcounter{equation}{0}

Inflation is the leading contender for the explanation of why the Universe is so big, old, and smooth \cite{guth,andrei,andrei2}. It 
gives a fully controlled prediction of the initial spectrum of almost scale invariant density fluctuations at superhorizon scales, which
are required to seed the cosmic structures observed in the late Universe \cite{perts}. These fluctuations are directly tested by the 
cosmic microwave background (CMB) measured by the WMAP and Planck experiments, and the fit to the inflationary predictions is excellent. Very recently the BICEP2 experiment claimed an observation of CMB polarization \cite{bicep} which fit the spectrum of primordial gravity waves \cite{stargw}, that can also be created during inflation. The BICEP2 results, if due to primordial gravity waves, point towards large field models of inflation, 
to explain the claimed large tensor-to-scalar ratio. Such models have large field excursions $\Delta \varphi > M_{Pl}$ during inflation, and the potential remains very flat and small in Planck units \cite{andrei2}. They are difficult to realize because at large field values the quantum corrections can be large. However, there are setups using a pseudo-Goldstone boson of some weakly broken symmetry as the inflaton 
\cite{natural}, whose approximate shift symmetry protects the potential from large corrections \cite{marco,monodromy,fluxmono}. In these constructions the inflaton's shift symmetry is a `phase rotation', and the inflaton is necessarily a pseudo-scalar (essentially a type of axion).

A natural question to ask, assuming that the BICEP2 results are verified, is whether it pins down the inflaton to be a pseudo-scalar. Here we argue that an alternative is to use a 
scalar Goldstone boson for a non-compact, spontaneously broken global scale symmetry, the dilaton, as the inflaton. This automatically accommodates large field variations since the symmetry and the vacuum manifold are non-compact. The underlying scale symmetry naturally gives rise to an effective shift symmetry of the dynamical inflaton, which 
protects its potential in exactly the same way as the axion shift symmetry. 
The relevant low-energy degrees of freedom are the metric $g_{\mu\nu}$, the dilaton field $\Phi$ and the matter degrees of freedom, containing the Standard Model, to which they couple. 
Scale invariance forbids a direct Einstein-Hilbert term in the action, so the leading operator controlling graviton dynamics is a dilaton-graviton coupling 
$\Phi^2 R$. The Planck scale arises from the dilaton VEV $\vev{\Phi} \sim M_{Pl}$. A fully scale invariant theory allows only a quartic dilaton self coupling, without a non-trivial minimum. Scale invariance will protect the inflaton potential from any loop corrections. 
An inclusion of small explicit breaking terms allows for a non-trivial dilaton VEV at large but finite values $O(M_{Pl})$ with a very flat potential. All corrections to the inflaton potential will be suppressed by the small parameters characterizing the sizes of the explicit breaking terms. 

In detail, the Einstein frame potential is a combination of exponentials of the form $e^{\epsilon \varphi/M_{Pl}}$, where $\varphi$ is the canonically normalized inflation and $\epsilon \ll 1$. The parameter $\epsilon$, which ensures that the potential is shallow, is controlled by the anomalous scaling dimensions of the explicit breaking terms.
Since the $\beta$-functions of classically marginal operators usually do not vanish at loop level, one generically expects the $\beta$-functions to provide small perturbative contributions 
to the parameter $\epsilon$ controlling the breaking.  Thus the explicit breaking of scale invariance they introduce is small and the potential remains flat and small even when these corrections are included.
This follows because the theory has a (non-linearly realized, but manifest) shift symmetry.
Hence, if the dilaton starts out far from its minimum (close to the origin in the original Jordan frame), the relevant Einstein frame field displacement is given by $\Delta \varphi \sim M_{Pl} \log (\Phi_0/M_{Pl})$ where $\Phi_0$ is the initial value of the dilaton. This displacement is easily larger than the Planck scale while within the regime of the effective theory.

A low-energy effective field theory of the inflaton as a dilaton is determined by a set of local operators arising from the spontaneous breaking of a scale invariant UV theory. Since $M_{Pl}$ must also emerge dynamically --- analogous to the pion decay constant --- due to the breaking of scale invariance, one may expect that the graviton dynamics is controlled by an operator in this expansion too. A standard route to realize this is induced gravity \cite{sakharov,adler}, wherein one starts with a dynamical spin-2 field coupled to matter, but where the gravitational coupling scale is determined by the renormalization of the spin-2 sector, that is radiative corrections from the matter fields generate the graviton kinetic term. After the scale symmetry is broken, these will control the coupling of the low-energy graviton, which remains massless because the diffeomorphism invariance of the theory remains unbroken. A more interesting --- and less well understood --- route is to imagine that the low-energy graviton is a composite of the underlying UV theory. The Weinberg-Witten theorem \cite{ww} prohibits starting with a local flat space field theory with a covariantly conserved stress-energy tensor and a \emph{massless}  composite graviton emerging from it and coupling to that stress-energy tensor. 
This nevertheless leaves open possibilities for generating a massless composite graviton. One can imagine either that the UV theory may not have a locally conserved stress-tensor, or that the low-energy matter fields --- which we observe ---  appear only in the low-energy effective stress-tensor that couples to the composite graviton. The latter possibility is analogous to what occurs in Seiberg duality, where the massless composite gauge bosons couple to IR degrees of freedom. In such a case, one can imagine that all known particles -- including the graviton --- are composites of some underlying theory. Given that the compositeness scale is high $\sim M_{Pl}$, there would be no
conflict with any experimental measurements.

This paper is organized as follows: in the next section we set up the effective theory, in section 3 we study different inflation potentials that can arise in theories with approximate scale invariance, while in section 4 we discuss the cutoff scale and higher order corrections. Section 5 contains comments on the effects of matter fields and their potential role on reheating. Finally we present our conclusions in section 6.

\section{The Setup}
\setcounter{equation}{0}

The main assumption we will make is that the low-energy, effective Lagrangian describing the gravity-inflaton system is scale invariant, with small explicit breaking terms responsible for setting the scale of symmetry breaking. Global scale transformations are given by $x^\mu \to \bar x^\mu= e^{-\lambda} x^\mu$, or equivalently $g_{\mu\nu}\to e^{-2\lambda} g_{\mu\nu}$. These have the effect $R\to e^{2\lambda} R$ on the scalar curvature, while generic operators transform as ${\cal O}\to e^{\lambda \Delta} {\cal O}$, where $\Delta$ is the scaling dimension of the operator. A Lagrangian is scale invariant if all operators have dimension 4.  The spontaneous breaking of such scale invariance is parameterized by the dilaton field $\Phi$, which is the Goldstone boson for broken scale invariance, and which will also serve as the inflaton in this setup. Once the dilaton is stabilized due to the presence of small explicit breaking terms, its VEV will give rise to the effective Planck scale. However, we will assume that at an initial time the dilaton is displaced far from its minimum, and the process of the dilaton rolling to its minimum will be the cause of inflation. 

The general scale invariant Lagrangian that we will be considering is given by
\begin{equation}
\Lag = \sqrt{-g} \left[ \tilde{\xi} \Phi^2 R  -\frac{1}{2} (\nabla \Phi )^2 - V(\Phi) \right] + \Delta \Lag(g_{\mu \nu},\Phi) + \Lag_M(g_{\mu \nu},\Phi,\Psi) \ .
\label{LagPhi}
\end{equation}
where $R$ is the Ricci scalar\footnote{We use the mostly plus signature $\eta_{\mu \nu} = \mathrm{diag}(-1,1,1,1)$, and $R^\alpha_{\ \beta \gamma \delta} = \partial_\gamma \Gamma^\alpha_{\beta \delta} + \Gamma^\alpha_{\lambda \gamma} \Gamma^\lambda_{\beta \delta} - (\gamma \leftrightarrow \delta)$, $R_{\mu \nu} = R^\alpha_{\ \mu \alpha \nu}$.} and $\tilde \xi$ is a dimensionless parameter.\footnote{The additional requirement of conformal invariance would fix the parameter $\tilde{\xi}=1/12$. We will not impose this assumption in this paper. Naive dimensional analysis (NDA) suggests $\tilde{\xi}= O(16 \pi^2)$.}  Note, that scale invariance forbids the presence of the usual Einstein-Hilbert term. 
The potential $V(\Phi)$ will be specified below, but exact scale invariance would require $V(\Phi) = \alpha^2 \Phi^4$, with a constant $\alpha$.  Scale invariance forbids large corrections to the dilaton potential, hence eliminating the $\eta$-problem. This remains valid even after including the loop corrections from the interactions with other fields, as long as these fields do not violate scale invariance explicitly. This will be the case if the masses of the fields interacting with the dilaton originate from the dilaton VEV itself, in which case the resulting corrections will just renormalize the coefficient of the $\Phi^4$ coupling. In the presence of small explicit breaking terms (which will be necessary to obtain a non-trivial VEV) the corrections to the dilaton potential will be suppressed by the small parameter characterizing the magnitude of the explicit breaking. 
As we will note below, this follows since the theory --- including the regulator --- has a manifest (non-linearly realized) shift symmetry, which arises from scale invariance after field redefinitions.
In turn this also guarantees that all the perturbative graviton loop corrections are completely under control, much like in the case of axion monodromy \cite{fluxmono} (see also \cite{rattaold}).

In order to recover Einstein gravity, the potential must give rise to a non-vanishing VEV for $\Phi$, $\vev{\Phi}^2 = M_{Pl}^2/2 \tilde \xi$.
$\Delta \Lag(g_{\mu \nu},\Phi)$ contains operators with extra derivatives and inverse powers of $\Phi$, for example the Weyl term involving $R^2$ would be in this part of the Lagrangian.  $\Lag_M(g_{\mu \nu},\Phi,\Psi)$ contains any other dynamics involving fields collectively denoted by $\Psi$ (such as the Standard Model (SM) fields), which may or may not be coupled to $\Phi$ (but they certainly couple to the metric in order to preserve Lorentz invariance). We will discuss the role of these two terms later.
We do not consider at this point any violation of Lorentz covariance, even if Lorentz invariance itself is only a symmetry of the low-energy and not of the underlying UV theory.

In order to understand the inflationary dynamics of this system, it is convenient to perform a Weyl transformation of the metric and go to the Einstein frame:
\begin{equation}
g_{\mu \nu} \to \Omega^2 g_{\mu \nu} \ ,
\end{equation}
where $\Omega = \Omega(x)$ satisfies
\begin{equation}
\Omega^2 \tilde \xi \Phi^2 = \frac{M_P^2}{2} \ .
\end{equation}
The rescaled Lagrangian is given by
\begin{equation}
\Lag = \sqrt{-g} \left[ \frac{M_{Pl}^2}{2} R - \frac{1}{2} (\nabla \varphi )^2 - V(\varphi) \right] + \Delta \Lag \! \left( \Omega^2(\varphi)g_{\mu \nu},\Phi(\varphi) \right) + \Lag_M \! \left( \Omega^2(\varphi)g_{\mu \nu},\Phi(\varphi), \Psi \right) \ ,
\label{LagVarphi}
\end{equation}
where
\begin{equation}
V(\varphi) = \frac{M_{Pl}^4}{4\tilde \xi^2 } \frac{V\!\left( \Phi(\varphi) \right)}{\Phi^4(\varphi)} \ .
\end{equation}
The relation between the original dilaton and the Einstein frame inflaton $\varphi$ is given by (given the boundary condition $\Phi(\varphi = 0) = \vev{\Phi}$)
\begin{equation}
\Phi(\varphi) = \vev{\Phi} \exp \left( \frac{\sqrt{\xi} \varphi }{M_{Pl}} \right) \ , \quad \frac{1}{\xi} = \frac{1}{2 \tilde{\xi}} + 6 \ .
\label{PhiVarphi}
\end{equation}
In this frame the original scale invariance of the theory will manifest itself in a shift symmetry for the inflaton 
\begin{equation}
\varphi \to \bar \varphi = \varphi + \frac{M_{Pl}}{\sqrt{\xi}} \lambda \  .
\label{shift}
\end{equation} 
Thus \eq{LagVarphi} can be thought of as the non-linearly realized Lagrangian for the spontaneously broken non-compact group of scale transformations, where the above shift symmetry is the remnant of the original scale invariance. The Einstein-Hilbert term is shift symmetric, since it does not contain $\varphi$. The kinetic term for the scalar is shift symmetric because it contains only derivatives. The scalar potential term $V(\varphi )$ becomes a constant (if we started out with a quartic $\Phi^4$ in the Jordan frame, as required in the absence of explicit breaking terms). The terms in $\Delta \Lag$ already contain derivatives of $\varphi$ only, and thus will be obviously shift invariant. The only non-trivial terms are those that involve matter fields coupled to $\varphi $ in  ${\cal L}_M$: here explicit powers of $e^{\sqrt{\xi} \varphi/M_{Pl}}$ will appear from the Weyl transformation of the metric, seemingly giving rise to non-derivative interactions. The important point is that such factors will also be present in the kinetic terms of the matter fields: once the matter fields are suitably redefined in order to canonically normalize their kinetic terms, the inflaton will again appear only derivatively coupled, obeying the shift symmetry. Hence all the terms in Eq. (\ref{LagVarphi}) which were originally exactly scale invariant remain invariant under the shift symmetry.

Notice also that, given $\varphi = (M_{Pl}/\sqrt \xi) \log (\Phi/\vev{\Phi})$, if the dilaton field starts out at small values $\Phi_0 \sim 0$ far from the minimum of the potential and moves out to $\vev{\Phi} \sim M_{Pl}$, the field space range for $\varphi$ can be larger than $M_{Pl}$ without ever leaving the regime of validity of the effective theory.  For example assuming $\Phi_0  \sim 10^{-15} \vev{\Phi} \sim \TeV$, we find $|\Delta \varphi |\sim  15 M_{Pl}$, a seemingly super-Planckian field excursion in the Einstein frame.\footnote{If $\Phi_0$ is orders of magnitude larger than its VEV, $\Delta \varphi$ is also super-Planckian. Such a large field excursion is a direct consequence of using the Goldstone field for a non-compact symmetry group.}

The scale invariant $\alpha^2 \Phi^4$ dilaton potential, yields a completely flat constant potential independent of $\varphi$ in the Einstein frame. This is again a simple consequence of the shift symmetry \eq{shift}.
Only derivative couplings of $\varphi$ are allowed, which are contained in $\Delta \Lag$ in \eq{LagVarphi}.
However for a completely flat potential the VEV $\vev{\Phi}$ (and the Planck scale) remain undetermined. One needs to systematically incorporate small explicit breaking terms into the Lagrangian which can easily fix the dilaton VEV at large values. Such explicit breaking terms  could possibly originate from the interactions with additional matter contained in $\Lag_M$, in particular they could potentially be due to interactions with the SM fields. As long as the explicit breaking induced by these terms is weak, the shift symmetry \eq{shift} will remain approximately valid, and will continue to protect the low energy theory \eq{LagVarphi} from large corrections. In the next section we will consider three well-motivated simple forms of the explicit breaking terms. 

\section{Approximately Scale Invariant Inflaton Potentials}
\setcounter{equation}{0}

The goal of this section is to present examples of well-motivated potentials that systematically incorporate small explicit breakings of scale invariance and examine their experimental consequences. In every example we will require that the cosmological constant vanishes at the minimum of the potential. This is needed for two reasons: inflation will not end if the minimum of the potential is not small, and the observed cosmological constant is very small (in Planck units). Requiring scale invariance does not in itself say anything about the cosmological constant: as we saw before, scale invariance allows a $\Phi^4$ potential for the dilaton, which after moving to the Einstein frame is a contribution to the cosmological constant. This term can be understood as the vacuum energy created during the phase transition from the scale invariant phase to the spontaneously broken one considered here.\footnote{For recent discussions on attempts to use weakly broken scale invariance to reduce the cosmological constant see~\cite{CPRtypes}.}

\subsection{A Single Relevant Operator\label{sec:GW}}

The first example takes the effect of a single marginally relevant operator with dimension $4-\epsilon$ into account, while also requiring the cosmological constant to vanish at the minimum of the potential. This type of potential~\cite{RZ,Higgslike} naturally shows up in warped extra dimensions~\cite{RS1} after modulus stabilization via the Goldberger-Wise mechanism~\cite{GW1,GW2} (which indeed corresponds to turning on a marginally relevant operator in the dual conformal field theory language). The resulting approximately scale invariant potential is 
\begin{equation}
V(\Phi) = \Phi^4 \left( \alpha + \beta \Phi^{-\epsilon} \right)^2 \ ,
\end{equation} 
where $\epsilon$ corresponds to the anomalous dimension of the operator breaking scale invariance, $\epsilon \ll 1$. This potential is minimized at
\begin{equation}
\vev{\Phi} = \left( \frac{\alpha}{\beta} \right)^{1/\epsilon} \ ,
\label{VEV1}
\end{equation}
and the potential at the minimum vanishes to reproduce an (approximately) zero vacuum energy density at the end of inflation. This is where we tune away the cosmological constant (at this point neglecting any other contributions to the cosmological constant from $\Lag_M$).

The inflaton potential in the Einstein frame reads
\begin{equation}
V(\varphi) = \frac{M_{Pl}^4}{4} \frac{\alpha^2}{\tilde \xi^2} \left( 1 - e^{-\epsilon \sqrt \xi \varphi/M_{Pl}} \right)^2 \ .
\label{potential}
\end{equation}
As expected this is a very flat potential, as long as $\epsilon \ll 1$. Flatness is a result of a small explicit breaking of scale invariance. Note that the form of the potential in the Einstein frame is the same as that of the Starobinsky model~\cite{Starobinsky}, with the important difference that the exponent here is controlled by the amount of explicit breaking in the field theory. In contrast, in the original Starobinsky model the exponent is fixed by $4D$ general covariance. To understand why the Starobinsky potential is a special case of \eq{potential}, however, all one needs is scaling symmetry. The starting action of \cite{Starobinsky} can be thought of as a special case of scale invariant theory where the breaking of scale invariance is induced purely gravitationally, by an explicit $M_{Pl}^2 R$ term. This immediately explains the necessity that in the Starobinsky inflation, the $R^2$ term {\it must} dominate over $M_{Pl}^2 R$ to yield inflation: the scale symmetry breaking term {\it must} be subleading in the UV for the protection mechanism to be operational. This is also the reason behind the emergence of the same type of potentials in the context of induced gravity, as explained in \cite{ksy}.

The slow-roll parameters are given by
\begin{equation}
\epsilon_V = \frac{M_{Pl}^2}{2} \left( \frac{V'(\varphi)}{V(\varphi)} \right)^2 = \frac{2 \epsilon^2 \xi }{( 1 - e^{\epsilon \sqrt \xi \varphi/M_{Pl}})^{2}} \ ,
\qquad 
\eta_V = M_{Pl}^2 \frac{V''(\varphi)}{V(\varphi)} = \epsilon_V \left( 2 - e^{\epsilon \sqrt \xi \varphi/M_{Pl}} \right) \ .
\end{equation}
The number of e-folds of inflation is well approximated by
\begin{equation}
N \simeq \frac{1}{M_{Pl}^2} \int_0^{\varphi_0} \frac{V'(\varphi)}{V(\varphi)} d \varphi = \frac{1}{2 \epsilon^2 \xi} \left[ \left( e^{\epsilon \sqrt \xi \varphi_0/M_{Pl}} -1 \right) - \frac{\varphi_0}{\sqrt{2}M_{Pl}} \right] \ .
\end{equation}
Notice that all the above expressions depend only on the combination $\epsilon \sqrt \xi$, which is the single parameter needed to characterize this model.\\

\begin{figure}[!t]
\begin{center}
\includegraphics[width=3in]{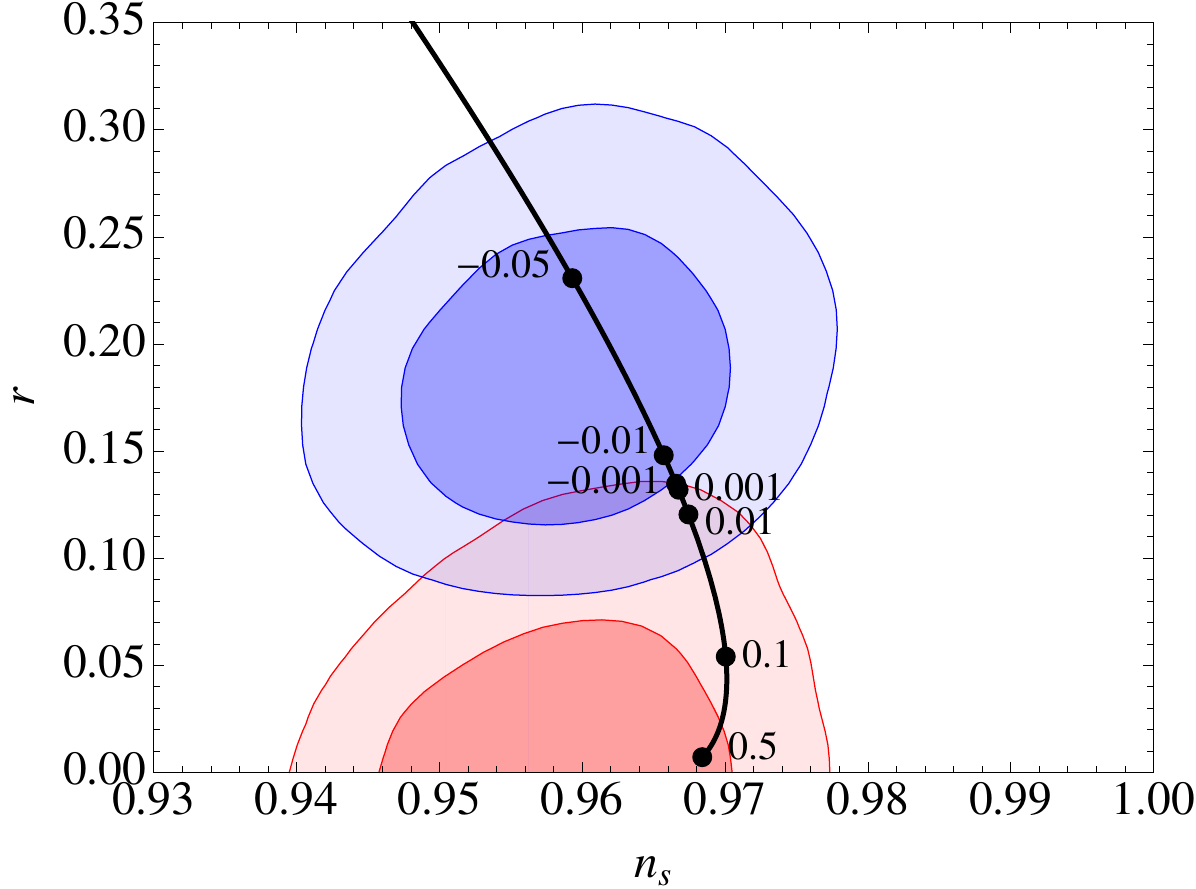}
\caption{Values of $n_s$ and $r$ for $\epsilon \sqrt \xi \in [-0.5,0.5]$, for $\varphi_0 < \vev{\varphi} = 0$. The same results are obtained for $\varphi_0 > \vev{\varphi} = 0$, but with opposite signs for $\epsilon$.
The points shown correspond to $\epsilon \sqrt \xi = -0.001, -0.01, -0.05, 0.001, 0.01, 0.1, 0,5$.}
\label{rns1}
\end{center}
\end{figure}

With these results we can compute the scalar power spectrum $\mathcal{P}_s$, the tensor-to-scalar ratio $r$, and the tilt of the primordial scalar perturbations $n_s$, at CMB horizon exit.
These are given by
\begin{equation}
\mathcal{P}_s \simeq \frac{1}{24 \pi^2} \frac{V^\star}{M_{Pl}^4} \frac{1}{\epsilon_V^\star} \ , \qquad
\end{equation}
\begin{equation}
r = \frac{\mathcal{P}_t}{\mathcal{P}_s} \simeq 16 \epsilon_V^\star \ , \qquad \qquad n_s = 1+ \frac{d \ln \mathcal{P}_s}{d \ln k} \simeq 1 + 2 \eta_V^\star - 6 \epsilon_V^\star \ ,
\end{equation}
where all the $\star$ parameters are evaluated at $\varphi = \varphi_{cmb}$ such that $N(\varphi_{cmb}) \simeq 60$.
We show in \fig{rns1} the values of $n_s$ and $r$ while varying $\epsilon \sqrt \xi \in [-0.5,0.5]$, for $\varphi_0 < \vev{\varphi} = 0$, corresponding to almost marginal perturbations. The same results are obtained for $\varphi_0 > \vev{\varphi} = 0$, but with opposite signs for $\epsilon$.
The points shown correspond to $\epsilon \sqrt \xi = -0.001, -0.01, -0.05, 0.001, 0.01, 0.1, 0,5$. If we insist on solutions with $\varphi_0 < \vev{\varphi}$, we can see that this model can accommodate both very small values of $r$ (for relatively large anomalous dimensions $\epsilon \sqrt \xi \sim O (0.1)$, that is marginally relevant perturbations), while $r$ can be pushed into the region favored by BICEP2 for $\epsilon \sqrt \xi < 0$, corresponding to marginally irrelevant perturbations. Similar observations were noted in specific 
constructions in \cite{sugra}. Again, as we see here the underlying reason for it is just small explicit breaking of scaling symmetry.

The COBE normalization of the scalar power spectrum, measured to be $(\mathcal{P}_s)_{exp} \sim 10^{-9}$, enforces a constraint on the parameter $\alpha$ in the potential, for fixed $\epsilon$ and $\tilde \xi$. Explicitly one obtains
\begin{equation}
\mathcal{P}_s = \frac{\alpha^2}{24 \pi^2 \tilde \xi^2 } \frac{\sinh^4 (\epsilon \sqrt{\xi} \varphi_{cmb}/2M_{Pl})}{\epsilon^2 \xi} \ ,
\end{equation}
where $\varphi_{cmb}$ is a function of $\epsilon \sqrt{\xi}$.
Since $\mathcal{P}_s$ increases with $\epsilon \sqrt{\xi}$, smaller values of the explicit breaking parameter $\epsilon$ --- and therefore better slow-roll approximation --- accommodate the observed power spectrum more easily, as one would expect from general inflationary phenomenology. However, from \eq{VEV1} one naturally expects that $\epsilon$ is of the order of $1/\ln(M_{Pl}/\Lambda_\epsilon)$, where the scale $\Lambda_\epsilon$ parametrizes the onset of scaling symmetry breaking. For instance $\Lambda_\epsilon \sim 10^{\pm 3} M_{Pl}$ yields $\epsilon \sim 0.1$, while $\Lambda_\epsilon \sim 10^{\pm 17} M_{Pl}$ gives $\epsilon \sim 0.01$.
Taking $\epsilon \sqrt \xi = \pm 0.01$ and for the most favorable case of $\tilde \xi \simeq 16 \pi^2$ (notice that $\mathcal{P}_s$ decreases with increasing $\tilde \xi$), the normalization of the scalar power spectrum is given by 
\begin{equation}
\mathcal{P}_s \simeq \left( \frac{\alpha}{0.1} \right)^2 \times 10^{-9} \ ,
\end{equation}
which requires a perturbative value of $\alpha$ compared to its NDA estimate $\alpha \sim 4 \pi$.

\subsection{A Cosh Potential}

Another simple potential could arise in the presence of a marginally relevant and marginally irrelevant perturbation. For simplicity we take their dimensions to be $4\pm \epsilon$, though one could of course also choose two independent dimensions. 
\begin{equation}
V(\Phi) = - \alpha^2 \Phi^4 + \beta^2 \Phi^{4-\epsilon} + \gamma^2 \Phi^{4+\epsilon} \ .
\end{equation} 
The minimum of the potential is at
\begin{equation}
\vev{\Phi} = \left( \frac{2 \alpha^2 + \sqrt{4 \alpha^4 + \beta^2 \gamma^2 (4 - \epsilon) (4 + \epsilon)}}{\gamma^2 (4 + \epsilon)} \right)^{1/\epsilon} \, .
\end{equation}
The inflaton potential in the Einstein frame reads, after fixing $\beta$ again such that $V(\vev{\Phi}) = 0$,
\begin{equation}
V(\varphi) = \frac{M_{Pl}^4}{4} \frac{\alpha^2}{\tilde \xi^2} \left( \cosh(\epsilon \sqrt \xi \varphi/M_{Pl}) - 1 \right) \ .
\end{equation}
This potential is clearly the non-compact analogue of the generic axion-type potentials for the case of a broken compact symmetry. Note, that the analogue of the axion decay constant appearing here is effectively given by $M_{Pl}/\epsilon \sqrt{\xi}$, which can be $\gg M_{Pl}$ for small $\epsilon$. However, here obtaining a `large decay constant' and allowing for an even larger range of variation of $\varphi$ is straightforward.

\begin{figure}[!t]
\begin{center}
\includegraphics[width=3in]{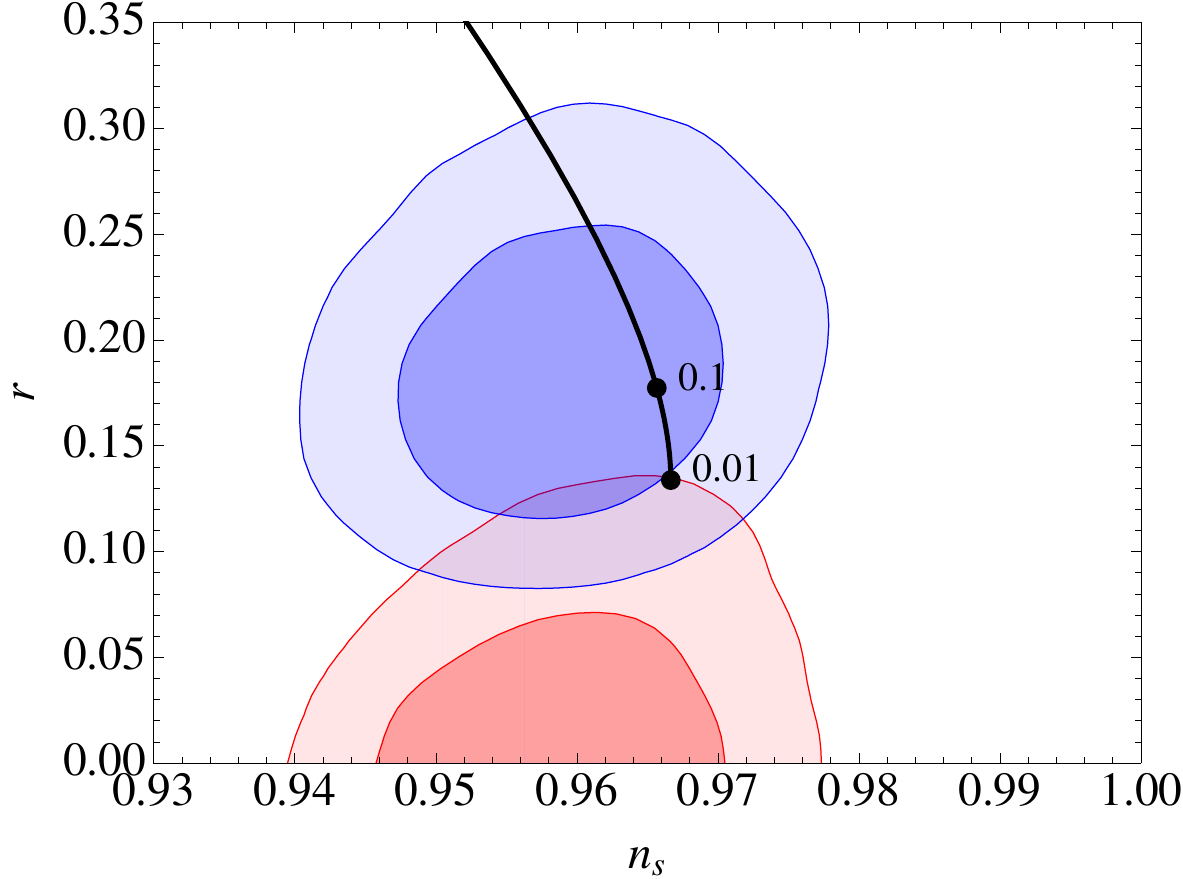}
\caption{Line of values of $n_s$ and $r$ for $\epsilon \sqrt \xi \in (0, 0.5]$, with points at $\epsilon \sqrt \xi = 0.1, 0.01$, for either sign of $\varphi_0$. The same results are obtained for negative $\epsilon$.}
\label{rns2}
\end{center}
\end{figure}

In this case the slow-roll parameters and the number of e-folds of inflation are
\begin{equation}
\epsilon_V = \frac{1}{2} \epsilon^2 \xi \coth^2(\epsilon \sqrt \xi \varphi/2M_{Pl}) \ , \qquad \qquad
\eta_V = \frac{\epsilon_V}{\cosh(\epsilon \sqrt \xi \varphi/M_{Pl})} \ ,
\end{equation}
\begin{equation}
N \simeq \frac{2}{\epsilon^2 \xi} \log \left[ \cosh(\epsilon \sqrt \xi \varphi/2M_{Pl}) \right] \ ,
\end{equation}
which again only depend on the combination $\epsilon \sqrt \xi$. In \fig{rns2} we show the line of values of $n_s$ and $r$ for $\epsilon \sqrt \xi \in (0, 0.5]$, with points at $\epsilon \sqrt \xi = 0.1, 0.01$, for either sign of $\varphi_0$. The same results are obtained for negative $\epsilon$. Small values of $|\epsilon|$ yield approximately the same result as for $\epsilon = 0.01$ (which is also very similar to the result at small $\epsilon$ for the previous potential, see Fig.~\ref{rns1}). Thus this particular model predicts a relatively large tensor-to-scalar ratio $r\gsim 0.1$. This is not surprising since the potential is an extrapolation of the quadratic potential, which generically yields larger $r$ \cite{andrei2,fluxmono}.

The normalization of the scalar power spectrum reads,
\begin{equation}
\mathcal{P}_s = \frac{\alpha^2}{12 \pi^2 \tilde \xi^2 } \frac{\sinh^2 (\epsilon^2 \xi N_{cmb})}{\epsilon^2 \xi} \ ,
\end{equation}
again an increasing function of $\epsilon \sqrt{\xi}$ and decreasing with $\tilde \xi$.
Taking $\tilde \xi \simeq 16\pi^2$ and for $\epsilon \sqrt \xi = \pm 0.01$,
\begin{equation}
\mathcal{P}_s \simeq \left( \frac{\alpha}{0.1} \right)^2 \times 10^{-9} \ .
\end{equation}

\subsection{A Potential with a Matter Induced Cosmological Constant}

Let us now assume that matter interactions contained in $\Lag_M$ generate a non-zero contribution to the vacuum energy.
In addition we include the effect of a marginally relevant operator with dimension $4-\epsilon$. 
The resulting potential is
\begin{equation}
V(\Phi) = \alpha^2 \Phi^4 - \beta^2 \Phi^{4-\epsilon} + \Lambda_M^4 \ .
\end{equation} 
which has its minimum at
\begin{equation}
\vev{\Phi} = \left( \frac{\beta^2 (4 - \epsilon)}{4 \alpha^2} \right)^{1/\epsilon} \ ,
\end{equation}
The inflaton potential in the Einstein frame is, after fixing $\Lambda_M$ such that the overall cosmological constant vanishes at the minimum $V(\vev{\Phi}) = 0$,
\begin{equation}
V(\varphi) = \frac{M_{Pl}^4}{4(4-\epsilon)} \frac{\alpha^2}{\tilde \xi^2} \left[ 4 \left( 1 - e^{-\epsilon \sqrt \xi \varphi/M_{Pl}} \right) - \epsilon \left( 1 - e^{-4 \sqrt \xi \varphi/M_{Pl}} \right) \right] \ .
\label{potexp}
\end{equation}
This is an example of racetrack inflation (see for example \cite{fernando}).
This potential gives rise to a long slow-roll inflation only for $\varphi > \vev{\varphi} = 0$, given that for negative values of $\varphi$ the constant $\Lambda_M^4$ term in the dilaton potential 
dominates.\footnote{Depending on the value of $\xi$, power law inflation may be possible for $\varphi < 0$, but we will ignore this case here.} In addition $\epsilon > 0$ is needed with all other signs fixed; if 
$\epsilon < 0$, one needs to change the signs of the potential terms according to $\alpha^2 \to - \alpha^2$, $\beta^2 \to - \beta^2$, $\Lambda_M^4 \to - \Lambda_M^4$.

In this case the slow-roll parameters are
\begin{equation}
\epsilon_V = \frac{ \epsilon^2 \xi \left( 1- e^{(\epsilon-4) \sqrt \xi \varphi/M_{Pl}} \right)^2}{2\left( 1 - \frac{4-\epsilon}{4} e^{\epsilon \sqrt \xi \varphi/M_{Pl}} - \frac{\epsilon}{4} e^{(\epsilon-4) \sqrt \xi \varphi/M_{Pl}} \right)^2} \ ,
~~~ \eta_V =  \frac{\epsilon \xi(\epsilon - 4 e^{(\epsilon-4) \sqrt \xi \varphi/M_{Pl}})}{1 - \frac{4-\epsilon}{4} e^{\epsilon \sqrt \xi \varphi/M_{Pl}} - \frac{\epsilon}{4} e^{(\epsilon-4) \sqrt \xi \varphi/M_{Pl}}} \ .
\end{equation}
The number of e-folds can be computed analytically leading to an expression involving hypergeometric functions. However, in order to get an idea of the parametric dependence of the predictions of the model on its parameters we can expand the potential \eq{potexp} in a Taylor series in the exponentials when $\varphi > M_{Pl}$, which is clearly required to be in the slow-roll regime. Then (also neglecting terms proportional to $\epsilon$ where $\epsilon$ is not in the exponential),
\begin{equation}
V(\varphi) = M_{Pl}^4 \frac{\alpha^2}{\tilde \xi^2} \left[ 1 - e^{-\epsilon \sqrt \xi \varphi/M_{Pl}} \right] \ .
\label{potexpexp}
\end{equation}
The slow-roll parameters reduce to 
\begin{equation}
\epsilon_V = \frac{ \epsilon^2 \xi}{2\left( 1 - e^{\epsilon \sqrt \xi \varphi/M_{Pl}} \right)^2} \ ,
\qquad \qquad  \eta_V =  \frac{\epsilon^2 \xi}{1 - e^{\epsilon \sqrt \xi \varphi/M_{Pl}}} \ .
\end{equation}

\begin{figure}[!t]
\begin{center}
\includegraphics[width=3in]{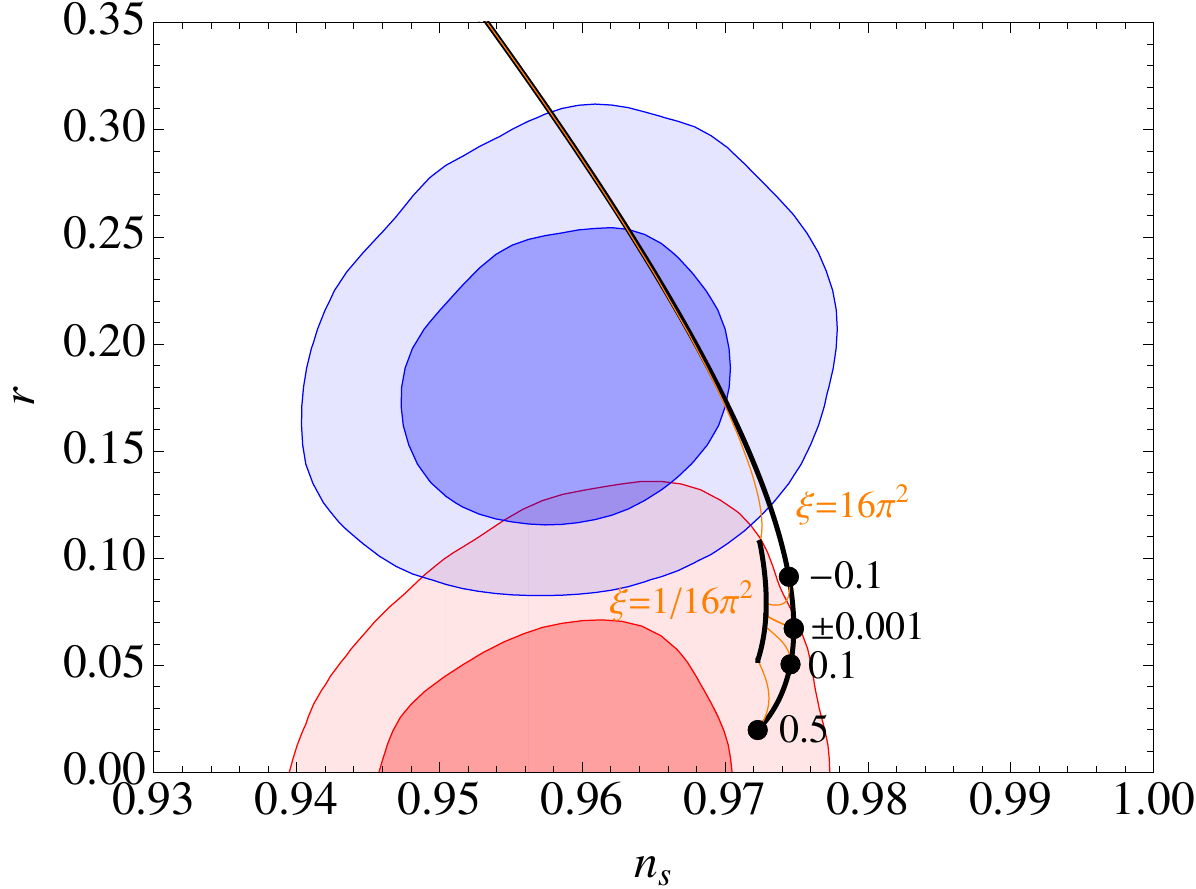}
\caption{Values of $n_s$ and $r$ for $\epsilon \in [-0.5,0.5]$ and $\tilde \xi \in [1/16 \pi^2, 16 \pi^2]$. The dots correspond to $\epsilon \sqrt \xi = -0.1, -0.001, 0.001, 0.1, 0.5$ and $\tilde \xi = 16 \pi^2$, with orange lines of constant $\epsilon$ and varying $\tilde \xi$ up to $\tilde \xi = 1/16 \pi^2$. The black lines are for fixed $\tilde \xi = 1/16 \pi^2, 16 \pi^2$ and varying $\epsilon$.}
\label{rns3}
\end{center}
\end{figure}

The results of $n_s$ and $r$ are shown in \fig{rns3}, for $\epsilon \in [-0.5,0.5]$ and $\tilde \xi \in [1/16 \pi^2, 16 \pi^2]$. The dots correspond to $\epsilon \sqrt \xi = -0.1, -0.001, 0.001, 0.1, 0.5$ and $\tilde \xi = 16 \pi^2$, with orange lines of constant $\epsilon$ and varying $\tilde \xi$ up to $\tilde \xi = 1/16 \pi^2$. The black lines are for fixed $\tilde \xi = 1/16 \pi^2, 16 \pi^2$ and varying $\epsilon$.
Notice that the sensitivity on $\epsilon$ decreases for decreasing $\tilde \xi$.
This follows because as $\tilde \xi$ becomes small, by \eq{PhiVarphi}, 
$\xi \simeq 2\tilde \xi$ is decreasing as well. 
Hence the slow-roll parameters in this limit are getting smaller and the slow-roll approximation is progressively more efficient.
Thus larger variation of $\epsilon$ will be allowed without spoiling the slow-roll approximation.
The most favorable limit in this case is  $\tilde \xi \simeq 16 \pi^2$, and so the COBE normalization of the scalar power spectrum, taking for example $\epsilon \sqrt \xi = \pm 0.01$, is
\begin{equation}
\mathcal{P}_s \simeq \left( \frac{\alpha}{0.05} \right)^2 \times 10^{-9} \ .
\end{equation}

\subsection{Power Law Potentials}

As a final example, let us switch our vantage point and consider other plausible potentials directly in the Einstein frame, thinking about the dilaton dependent terms in terms of the canonically normalized Goldstone inflaton. This viewpoint opens up a panorama of scaling symmetry breaking potentials which arise from loop-generated terms. By the original assumption that the breaking of scaling symmetry is small, these terms will be naturally small also, remaining under the protection of the shift symmetry. 

Specifically, one can imagine radiatively generated contributions to the dilaton/inflaton potential from say, $n$-loop diagrams in perturbation theory. Scaling symmetry sets the functional form to be $\sim \beta_n \Phi^4 [\log(\Phi/\Lambda_{\epsilon})]^n$, where $\beta_n$ is proportional to the $n$-loop $\beta$-function. In the Einstein frame these give rise to power law potentials 
\begin{equation}
V_n \sim \beta_n \vev{\Phi}^4  \left(\frac{\sqrt{\xi} \varphi}{M_{Pl}} \right)^n \sim \beta_n \frac{\varphi^n}{M_{Pl}^{n-4}} \ , 
\label{powerlaw}
\end{equation}
where the shift symmetry of $\varphi$ ensures that $\beta_n$ is small and under control, with the expectation $\beta_n \sim O(1/16\pi^2)^n$. 

Such power law potentials could be the leading terms driving inflation, depending on the details of the explicit scale invariance breaking sources (most likely those generated at $(n=1)$-loop).
Even if they are not the leading terms, if the effective scale of the potential \eq{powerlaw}, $\beta_n \vev{\Phi}^4$, is close to the scale of inflation, since the inflaton is automatically normalized by $M_{Pl}$, such terms could yield interesting corrections to the leading order inflationary potential that could leave their fingerprints on the sky (see, e.g.\ the discussion in the last reference of 
\cite{fluxmono} for similar corrections and their imprints on the sky in pseudoscalar-driven inflation). The precise determination of such phenomena is beyond the scope of the present work.

\section{Cutoff Scale and Higher Order Corrections}
\setcounter{equation}{0}

Here we address the regime of validity of our effective field theory. This task is best performed in the Einstein frame \eq{LagVarphi}.
There we can identify the inflaton decay constant, associated to the spontaneous breaking of scale invariance, as
\begin{equation}
f = \frac{M_{Pl}}{\sqrt{\xi}} \ .
\end{equation}
The cutoff of such an effective theory lies at or below
\begin{equation}
\Lambda_{UV} = \frac{4 \pi}{\sqrt{\xi}} M_{Pl} \ .
\end{equation}
We can explicitly check that this is the case by studying the operators at higher order in derivatives encoded in $\Delta \Lag$ in \eq{LagPhi}, and identifying the effective cutoff scale that suppresses them.
One such term is $R^2$, which in the Einstein frame gives rise to
\begin{equation}
\frac{1}{g_R^2} R^2 \to \frac{1}{g_R^2} \left[ R + 6 \left( \frac{\sqrt{\xi}}{M_{Pl}} \nabla^2 \varphi - \frac{\xi}{M_{Pl}^2} (\nabla \varphi )^2\right) \right]^2 \ .
\end{equation}
Each of the terms on the r.h.s.\ indicates that the cutoff lies at, or somewhat below, $\Lambda_{UV}$. For instance, the $R^2$ term can be regarded as arising from integrating out a scalar of mass $M_R^2 \simeq g_R^2 M_{Pl}^2$, which for the NDA estimate $g_R \sim 4 \pi$, sets the cutoff at $\Lambda_{UV} \approx M_R \sim 4 \pi M_{Pl}$.
Similarly, the other two terms set the cutoff at $\Lambda_{UV} \approx (g_R^2/\xi) M_{Pl} \sim 4 \pi M_{Pl}$.
Notice however that by taking small values of $\tilde \xi$ in \eq{PhiVarphi}, for which $\xi \simeq 2 \tilde \xi$, this latter cutoff can be raised above the naive expectation, contrary to the $R^2$ case.
The same behavior as for $R^2$ is found for the $R_{\mu \nu}^2/\tilde g_R^2$ operator. In this case it corresponds to a spin-2 ghost field with mass $\tilde M_{R}^2 \simeq \tilde g_R^2 M_{Pl}^2$. As long as $\tilde g_R^2$ is sufficiently large, the cutoff is above $M_{Pl}$.

Another example of operator in $\Delta \Lag$ is,
\begin{equation}
\frac{1}{g_{\Phi}^4} \frac{\left[ (\nabla \Phi )^2 \right]^2}{\Phi^4} \to \frac{1}{g_{\Phi}^4} \frac{\xi^2}{M_{Pl}^2} \left[ (\nabla \varphi )^2 \right]^2 \ ,
\end{equation}
in the Jordan and Einstein frames.
This is again a Planck-suppressed operator, and for $g_\Phi \sim 4 \pi$ the cutoff actually lies at $\Lambda_{UV} \sim 4 \pi M_{Pl}$.
All of the scale invariant higher dimensional corrections in $\Delta \Lag$ will share this property.\footnote{One should keep in mind in this discussion that loop corrections to the graviton sector point in fact to an effective low-energy cutoff somewhat {\it below} $M_{Pl}$. Generically, because of a number $N$ of light matter modes, the low-energy effective action begins to break down at scales $\Lambda_{UV,grav} \sim 4 \pi M_{Pl}/\sqrt{N}$. Therefore, if $M_{Pl}$ is fixed before inflation and never changes, this implies an upper bound on the number of light matter degrees of freedom, $N \lesssim (M_{Pl}/H)^2$, in order to guarantee that the geometry of inflation looks $4D$ for a Hubble scale $H$ during inflation. If BICEP2 is correct, $H \sim 10^{14} \GeV$ and thus $N \lesssim 10^8$, which constitutes a rather mild constraint. Therefore, these arguments do not affect the dynamics of the inflaton, for the reasons we discussed above.}

We should stress that the inflaton is derivatively coupled, that is it only appears through its derivatives $\nabla \varphi$, in any of the operators in $\Delta \Lag$.
Therefore, field excursion of the inflaton beyond $\Lambda_{UV}$ do not constitute a problem given that the inflaton potential ensures that its derivatives are small.
Large $\varphi$ values could be problematic in non-derivative terms, which are associated to the explicit breaking of the shift symmetry. However, for that very same reason --- as long as the breaking of the scaling/shift symmetry is weak --- they are kept small and under control, via $\epsilon$-suppression. This means, that even if the actual explicit breaking of scaling symmetry is below $M_{Pl}$ but is weak, the low energy theory remains extremely well protected by the approximate shift symmetry, essentially staying valid all the way up to the scale of quantum gravity, because the scaling symmetry breaking sector is very efficiently sequestered away from the low energy inflaton.


\section{Dynamics of Matter Fields and Reheating}
\setcounter{equation}{0}

Finally let us turn to the dynamics of the matter fields, which is clearly quite dependent on the UV completion.  We will assume that at very high energies the SM fields are still the proper degrees of freedom. If this is indeed the case, the couplings in the matter Lagrangian $\Lag_M$ are classically marginal (dimension 4) with the exception of the Higgs mass term. Thus at tree-level the SM Lagrangian is scale invariant, while the Higgs mass parameter constitutes a small explicit breaking of $O(m_H^2/M_{Pl}^2)$. A tree-level Higgs-dilaton quartic coupling is classically scale invariant and will thus not generate any mass for the dilaton: instead it contributes at loop-level to the dilaton quartic self interaction (after taking into account that the cutoff is proportional to the dilaton itself, a necessary condition to ensure that the UV regulator does not yield strong scaling symmetry breaking).\footnote{However such a Higgs-dilaton coupling would give rise to a large contribution to the Higgs mass, and for this reason should be small.} At loop-level the SM couplings run, but the $\beta$-functions at high energies are perturbatively small, being at most $O(1/16\pi^2)$. These effects will yield small explicit breaking parameters that could be potentially identified with the parameter $\beta_n$ in \eq{powerlaw}.

The exact form of the couplings between the dilaton/inflaton  and the SM matter fields are somewhat dependent on the details of the embedding of the SM fields into the scale invariant UV theory. To obtain their couplings, one can usually dress the dimensionful parameters, treated as spurionic fields, with the appropriate powers of $\Phi/\vev{\Phi} = e^{\varphi/f}$, with $f = M_{Pl}/\sqrt{\xi}$. Moreover, one must be aware that, even if absent in the Jordan frame, once in the Einstein frame (and for canonically normalized matter fields), derivative couplings of the inflaton to the SM fields are generated.

This is the case for instance for the coupling of the inflaton to the Higgs field:  a coupling of the form $-\sqrt{\xi} |H|^2 \partial^2 \varphi/M_{Pl}$ appears in the Einstein frame, through which the inflaton can decay to the longitudinal components of $WW$ and $ZZ$, and to the Higgs boson. This is consistent with the shift symmetry acting on the Goldstone $\varphi$, given that it is a  derivative coupling.
Notice furthermore that this decay $\varphi\to WW,ZZ,hh$ can only proceed if the shift symmetry is explicitly broken, which in this case comes in the form of $m_\varphi \neq 0$.
One might be concerned that, given that the Higgs field must get a VEV and break the electroweak symmetry, the associated Goldstone modes can be rotated away from the term above by going to the unitary gauge. However, one must notice that such a term induces a mixing between the Higgs boson and the inflaton, 
leading to a decay rate of (including the decay to the Higgs boson)
\begin{equation}
\Gamma_{\varphi \to WW,ZZ,hh} \simeq \frac{4 \xi}{32\pi} \frac{m_\varphi^3}{M_{Pl}^2} \simeq 0.5 \GeV \ \left( \frac{\xi}{1/12}\right) \left( \frac{m_\varphi}{10^{13} \GeV}\right)^3 \ ,
\end{equation}
where the reduced Planck mass is $M_{Pl} \simeq 2.5 \times 10^{18} \GeV$, and 
the mass of the inflaton, in the simplest example of Sec.~\ref{sec:GW} is given by 
\begin{equation}
\label{dilatonmass}
m_\varphi= M_{Pl} \frac{\alpha \epsilon \sqrt{\xi}}{\tilde\xi}
\simeq 10^{13} \left( \frac{\alpha}{0.1} \right) \left( \frac{\epsilon \sqrt{\xi}}{0.01}\right) \left( \frac{16\pi^2}{\tilde\xi} \right) \GeV \ .
\end{equation}

Although generically subleading in what regards the inflaton decay rate, we can also consider the coupling of the inflaton to two massless SM gauge bosons. It can be read off from the running of the corresponding gauge couplings, and is thus related to the associated $\beta$-functions. If the SM gauge sector arises as a composite or a partially composite of the sector that spontaneously breaks scale invariance, then the inflaton will couple to two gauge bosons proportionally to the change in the $\beta$-function of the SM group during the transition from the unbroken to the broken phase~\cite{Higgslike}:
\beq
\frac{\alpha_{SM}}{8 \pi} \left( b_{IR}-b_{UV}\right)  F^{\mu \nu}F_{\mu\nu} \sqrt{\xi} \frac{\varphi}{M_{Pl}} \ .
\label{eq:sigmaF}
\eeq
This coupling is loop suppressed compared to the $WW,ZZ$ couplings and is usually sub-leading (unless $\Delta b$ is very large), giving rise to a contribution to the dilaton width 
\beq
\Gamma_{\varphi\to 2g} \simeq \frac{\alpha^2_s}{256 \pi^3}\Delta b_s^2 \xi  \frac{m_\varphi^3}{M_{Pl}^2} \simeq 3 \keV \left( \frac{\xi}{1/12}\right) \left( \frac{m_\varphi}{10^{13} \GeV}\right)^3 (\Delta b_s)^2   \ ,
\eeq
where we have taken $\alpha_s \simeq 1/25$.

The reheat temperature is generically dominated by $\varphi \to WW,ZZ,hh$ decays, and is given by 
\beq
T_{RH}\sim g_*^{-1/4}(\Gamma M_{Pl})^{1/2} \sim 3 \times 10^8 \GeV \ , 
\eeq
for $g_*\sim O(100)$ and for the parameters chosen above. We can see that this temperature is high enough to accommodate (electroweak) baryogenesis, but sufficiently low to avoid restoration of high scale symmetries (like GUT) and prevent any regeneration of undesired topological defects.

\section{Summary}
\setcounter{equation}{0}

The class of models of inflation which we have built here rest on a few simple assumptions:
\begin{itemize}
\item We start with an 
underlying theory which includes gravity and has scaling symmetry. 
\item The scaling symmetry is spontaneously broken at a scale above the scale of inflation, simultaneously generating the low-energy Planck scale and allowing for the effective $4D$ cosmology with a set scale of inflation $V^{1/4} \sim \sqrt{M_{Pl} H}$. 
\item Explicit scaling symmetry breaking appears only through almost marginal operators with small anomalous dimensions (associated with non-trivial $\beta$-functions).  
\end{itemize}

Given these assumptions,\footnote{An issue we have sidestepped here regards the effects of the non-perturbative gravitational corrections, believed to be a serious obstacle to the existence of global symmetries in field theories coupled to gravity. 
In our case, since the global symmetry is scaling --- intricately related to the spacetime symmetries --- and the global symmetry breaking is very sensitive to the nature of the UV completion \cite{lenny}, it is not clear that the non-perturbative gravitational effects would be detrimental. If the UV theory has a full conformal symmetry linked to spacetime symmetries, such effects could be suppressed.}
we can write down the low-energy effective theory for the dilaton and the graviton, and their couplings to matter fields.
Since the dilaton is the Goldstone boson of the spontaneously broken
scaling symmetry, its dynamics has an approximate shift symmetry. Hence its potential, and specifically its mass, is suppressed by the factors of anomalous dimensions. 

Once we transform the theory to the Einstein frame, the dilaton naturally becomes a perfect candidate for an inflaton field. The non-compact nature of the symmetry allows for a range of field values larger than the Planck scale, and the approximate shift symmetry ensures that the slow-roll conditions are satisfied.
From the point of view of inflationary model building, this gives a natural and powerful tool to construct efficient models of large field inflation. This may be necessary to explain the observations if the claimed discovery of primordial tensor fluctuations by BICEP2 \cite{bicep} is correct. Our construction automatically accommodates a wide range of values of $r$ for a fixed spectrum of scalar perturbations $\mathcal{P}_s$. The simple examples we outlined allow for $r$ ranging easily from $0.2$ (for very small explicit breaking terms) all the way to small values of order $r\sim 0.01$ (for large explicit breaking).  Hence the mechanism provides a very broad class of models for observational tests.

The couplings of the dilaton to the SM fields are suppressed by anomalous dimensions or $\beta$-functions (or the Higgs mass over the Planck scale for the special case of the Higgs field). Nevertheless, the ensuing small couplings allow for a successful reheating. 

Thus our three assumptions 
lead to a viable model of inflation that can easily accommodate a scale invariant spectrum of perturbations and a significant ratio of tensor to scalar perturbations.

Furthermore, we believe that our formulation of the emerging inflaton from spontaneous scaling symmetry breaking gives a simple and straightforward framework to understand the origin of similar behavior reported in specific constructions, e.g. \cite{sugra}. From our point of view, the key reason for the existence of slow-roll chaotic inflation potentials is the approximate shift symmetry inherited from the scaling symmetry that controls the underlying dynamics, and the 
tensor-to-scalar ratio is a measure of its explicit breaking, being controlled by the anomalous dimension of the dominant scaling symmetry breaking operator (with small values of $r$ only attainable with large breaking). This also includes the original Starobinsky model, where from our perspective the scaling symmetry breaking is induced purely gravitationally.

In conclusion, we have found a simple set of rules required for a construction of models of large field inflation where the inflaton is a scalar. The dynamics is protected by an approximate shift symmetry, which is an avatar of the underlying scaling symmetry. This class of models should be a useful 
milestone complementing the large field pseudoscalar models \cite{monodromy,fluxmono} for the future observational tests of inflation.

\section*{Acknowledgments}
\setcounter{equation}{0}
\setcounter{footnote}{0}

We thank Albion Lawrence,  Liam McAllister, Alex Pomarol, Oriol Pujolas, Lorenzo Sorbo, and Alexander G. Westphal for useful discussions. We especially thank Paul McGuirk and John Stout for providing us the data files for the Planck and BICEP contour plots. C.C. thanks the Mainz Institute for Theoretical Physics (MITP) for its hospitality while this project was completed. J.S thanks the Theory Group at Boston University for its hospitality while this project the was completed. C.C. and J.S. are supported in part by the NSF grant PHY-1316222. N.K. and J.T. are supported in part by the DOE Grant DE-FG03-91ER40674.


\begin{thebibliography}{99}

\bibitem{guth} 
  A.~H.~Guth,
  Phys.\ Rev.\ D {\bf 23}, 347 (1981).

\bibitem{andrei}
A.~D.~Linde,
  Phys.\ Lett.\ B {\bf 108}, 389 (1982);
  A.~Albrecht and P.~J.~Steinhardt,
  Phys.\ Rev.\ Lett.\  {\bf 48}, 1220 (1982).

\bibitem{andrei2}
  A.~D.~Linde,
  Phys.\ Lett.\ B {\bf 129}, 177 (1983).
  
\bibitem{perts} 
  V.~F.~Mukhanov and G.~V.~Chibisov,
  JETP Lett.\  {\bf 33}, 532 (1981)
  [Pisma Zh.\ Eksp.\ Teor.\ Fiz.\  {\bf 33}, 549 (1981)];
  A.~H.~Guth and S.~Y.~Pi,
  Phys.\ Rev.\ Lett.\  {\bf 49}, 1110 (1982);
  J.~M.~Bardeen, P.~J.~Steinhardt and M.~S.~Turner,
  Phys.\ Rev.\ D {\bf 28}, 679 (1983);
  S.~W.~Hawking,
  Phys.\ Lett.\ B {\bf 115}, 295 (1982).
  
\bibitem{bicep} 
  P.~A.~R.~Ade {\it et al.}  [BICEP2 Collaboration],
  \arXiv{1403.3985}{astro-ph.CO}

\bibitem{stargw} 
  A.~A.~Starobinsky,
  JETP Lett.\  {\bf 30}, 682 (1979)
  [Pisma Zh.\ Eksp.\ Teor.\ Fiz.\  {\bf 30}, 719 (1979)].
       
\bibitem{natural} 
  K.~Freese, J.~A.~Frieman and A.~V.~Olinto,
  Phys.\ Rev.\ Lett.\  {\bf 65}, 3233 (1990);
  F.~C.~Adams, J.~R.~Bond, K.~Freese, J.~A.~Frieman and A.~V.~Olinto,
  Phys.\ Rev.\ D {\bf 47}, 426 (1993)
  \arXivold{hep-ph/9207245}.

\bibitem{nima} 
  N.~Arkani-Hamed, H.~-C.~Cheng, P.~Creminelli and L.~Randall,
  Phys.\ Rev.\ Lett.\  {\bf 90}, 221302 (2003)
  \arXivold{hep-th/0301218};
  N.~Arkani-Hamed, H.~-C.~Cheng, P.~Creminelli and L.~Randall,
  JCAP {\bf 0307}, 003 (2003)
  \arXivold{hep-th/0302034}.
      
\bibitem{marco} 
  J.~E.~Kim, H.~P.~Nilles and M.~Peloso,
  JCAP {\bf 0501}, 005 (2005)
  \arXivold{hep-ph/0409138}.
    
\bibitem{monodromy}
  E.~Silverstein and A.~Westphal,
  Phys.\ Rev.\ D {\bf 78}, 106003 (2008)
  \arXiv{0803.3085}{hep-th};
  L.~McAllister, E.~Silverstein and A.~Westphal,
  Phys.\ Rev.\ D {\bf 82}, 046003 (2010)
  \arXiv{0808.0706}{hep-th};
  L.~McAllister, E.~Silverstein, A.~Westphal and T.~Wrase,
  \arXiv{1405.3652}{hep-th}.
      
  \bibitem{fluxmono}   
  N.~Kaloper and L.~Sorbo,
  Phys.\ Rev.\ Lett.\  {\bf 102}, 121301 (2009)
  \arXiv{0811.1989}{hep-th};
  N.~Kaloper, A.~Lawrence and L.~Sorbo,
  JCAP {\bf 1103}, 023 (2011)
  \arXiv{1101.0026}{hep-th};
  N.~Kaloper and A.~Lawrence,
  \arXiv{1404.2912}{hep-th}.

\bibitem{sakharov}
  A.~D.~Sakharov,
  Sov.\ Phys.\ Dokl.\  {\bf 12}, 1040 (1968); 
  [Dokl.\ Akad.\ Nauk Ser.\ Fiz.\  {\bf 177}, 70 (1967)];
  [Gen.\ Rel.\ Grav.\  {\bf 32}, 365 (2000)].
  
  \bibitem{adler}
  A.~Zee,
  Phys.\ Rev.\ Lett.\  {\bf 42}, 417 (1979);
  S.~L.~Adler,
  Rev.\ Mod.\ Phys.\  {\bf 54}, 729 (1982)
  [Erratum-ibid.\  {\bf 55}, 837 (1983)].

\bibitem{stelle} 
  K.~S.~Stelle,
  Phys.\ Rev.\ D {\bf 16}, 953 (1977).
  
\bibitem{ww} 
  S.~Weinberg and E.~Witten,
  Phys.\ Lett.\ B {\bf 96}, 59 (1980).
  
\bibitem{rattaold} 
  R.~Contino, L.~Pilo, R.~Rattazzi and A.~Strumia,
  JHEP {\bf 0106}, 005 (2001)
  [hep-ph/0103104].
    
  
\bibitem{CPRtypes}
 B.~Bellazzini, C.~Csaki, J.~Hubisz, J.~Serra and J.~Terning,
  Eur.\ Phys.\ J.\ C {\bf 74}, 2790 (2014)
  \arXiv{1305.3919}{hep-th};
 F.~Coradeschi, P.~Lodone, D.~Pappadopulo, R.~Rattazzi and L.~Vitale,
  JHEP {\bf 1311}, 057 (2013)
  \arXiv{1306.4601}{hep-th}.

\bibitem{RZ}
 R.~Rattazzi and A.~Zaffaroni,
  JHEP {\bf 0104}, 021 (2001)
  \arXivold{hep-th/0012248}.


\bibitem{Higgslike}
B.~Bellazzini, C.~Cs\'aki, J.~Hubisz, J.~Serra and J.~Terning,
  Eur.\ Phys.\ J.\ C {\bf 73}, 2333 (2013)
  \arXiv{1209.3299}{hep-ph}.


\bibitem{RS1}
 L.~Randall and R.~Sundrum,
  Phys.\ Rev.\ Lett.\  {\bf 83}, 3370 (1999)
  \arXivold{hep-ph/9905221}.


\bibitem{GW1}
 W.~D.~Goldberger and M.~B.~Wise,
  Phys.\ Rev.\ Lett.\  {\bf 83}, 4922 (1999)
  \arXivold{hep-ph/9907447}.

\bibitem{GW2}
 C.~Csaki, M.~Graesser, L.~Randall and J.~Terning,
  Phys.\ Rev.\ D {\bf 62}, 045015 (2000)
  [hep-ph/9911406];
 W.~D.~Goldberger and M.~B.~Wise,
  Phys.\ Lett.\ B {\bf 475}, 275 (2000)
  [hep-ph/9911457];
C.~Csaki, M.~L.~Graesser and G.~D.~Kribs,
  Phys.\ Rev.\ D {\bf 63}, 065002 (2001)
  [hep-th/0008151].



\bibitem{Starobinsky}
 A.~A.~Starobinsky,
  Phys.\ Lett.\ B {\bf 91}, 99 (1980).
  
  
\bibitem{ksy} 
  N.~Kaloper, L.~Sorbo and J.~'i.~Yokoyama,
  Phys.\ Rev.\ D {\bf 78}, 043527 (2008)
  \arXiv{0803.3809}{hep-ph}.

\bibitem{sugra} 
R.~Kallosh and A.~Linde,
  JCAP {\bf 1307}, 002 (2013)
  [arXiv:1306.5220 [hep-th]];
  A.~Linde,
  arXiv:1402.0526 [hep-th];
  R.~Kallosh, A.~Linde and A.~Westphal,
  arXiv:1405.0270 [hep-th];
    J.~Ellis, M.~A.~G.~Garcia, D.~V.~Nanopoulos and K.~A.~Olive,
  arXiv:1405.0271 [hep-ph].


\bibitem{fernando} 
  J.~J.~Blanco-Pillado, C.~P.~Burgess, J.~M.~Cline, C.~Escoda, M.~Gomez-Reino, R.~Kallosh, A.~D.~Linde and F.~Quevedo,
  JHEP {\bf 0411}, 063 (2004)
  [hep-th/0406230].
  
\bibitem{rs2}
  L.~Randall and R.~Sundrum,
  Phys.\ Rev.\ Lett.\  {\bf 83}, 4690 (1999)
  [hep-th/9906064];
  S.~S.~Gubser,
  Phys.\ Rev.\ D {\bf 63}, 084017 (2001)
  [hep-th/9912001];
  N.~Arkani-Hamed, M.~Porrati and L.~Randall,
  JHEP {\bf 0108}, 017 (2001)
  [hep-th/0012148].
  
\bibitem{lenny} 
  R.~Kallosh, A.~D.~Linde, D.~A.~Linde and L.~Susskind,
  Phys.\ Rev.\ D {\bf 52}, 912 (1995)
  [hep-th/9502069].
  
\end{thebibliography}
\end{document}